\documentclass{article}
\usepackage[latin9]{inputenc}
\usepackage{amsmath}
\usepackage{amssymb}
\usepackage{graphicx}

\makeatletter
\usepackage{spconf}
\usepackage{booktabs}

\usepackage{pstricks, pst-node, pst-plot, pst-circ}
\usepackage{moredefs}

\usepackage{psfrag}

\usepackage{amsmath,graphicx}

\newcommand{\HP}{\text{H\hspace{-0.25mm}P}}
\newcommand{\LP}{\text{L\hspace{-0.25mm}P}}
\newcommand{\sigf}{f}

\newcommand{\blockbased}{block-based}
\newcommand{\meshbased}{mesh-based}

\newcommand{\MSE}{Mean Energy}

\newcommand{\spaceBelowFig}{\vspace{-4mm}}
\newcommand{\spaceBeforeLabel}{\vspace{-1mm}}

\newcommand{\fig}{Fig.}
\newcommand{\tab}{Tab.}

\newcommand{\Fig}[1]{\fig{}~#1}
\newcommand{\Tab}[1]{\tab{}~#1}

\newcommand{\Legall}{LeGall~5/3 wavelet}

\newcommand{\jpthreed}{JPEG~2000~3D}
\newcommand{\speckthreed}{SPECK3D}

\newcommand{\zweid}{\mbox{2-D}}
\newcommand{\dreid}{\mbox{3-D}}
\newcommand{\dreidt}{\mbox{3-D+t}}

\name{Wolfgang Schnurrer, Thomas Richter, Jürgen Seiler, Christian Herglotz, and André Kaup}
\address{Multimedia Communications and Signal Processing\\
Friedrich-Alexander-Universität Erlangen-Nürnberg (FAU), Cauerstr. 7, 91058 Erlangen, Germany\\
Email: \{ schnurrer, richter, seiler, herglotz, kaup \}@lnt.de}
\makeatother

\begin{document}
\ninept

\title{\dreid{} Mesh Compensated Wavelet Lifting for \dreidt{} Medical
CT Data}
\maketitle
\begin{abstract}
For scalable coding, a high quality of the lowpass band of a wavelet
transform is crucial when it is used as a downscaled version of the
original signal. However, blur and motion can lead to disturbing artifacts.
By incorporating feasible compensation methods directly into the wavelet
transform, the quality of the lowpass band can be improved. The displacement
in dynamic medical \dreidt{} volumes from Computed Tomography is
mainly given by expansion and compression of tissue over time and
can be modeled well by \meshbased{} methods. We extend a \zweid{}
\meshbased{} compensation method to three dimensions to obtain a
volume compensation method that can additionally compensate deforming
displacements in the third dimension. We show that a \dreid{} mesh
can obtain a higher quality of the lowpass band by 0.28~dB with less
than 40\% of the model parameters of a comparable \zweid{} mesh.
 Results from lossless coding with \jpthreed{} and \speckthreed{}
show that the compensated subbands using a \dreid{} mesh need about
6\% less data compared to using a \zweid{} mesh.
\end{abstract}

\section{Introduction}

\label{sec:intro}

Dynamical \dreidt{} volumes can become very large in terms of storage
so a scalable representation is desirable, e.g., for an efficient
access and transmission. A downscaled version allows faster access,
e.g., for browsing purposes. The lowpass band of a wavelet transform
can be used as such a representation. By incorporating a compensation
method directly into the wavelet transform, the transform can be adapted
to the signal. A compensated wavelet transform in temporal direction
is known as Motion Compensated Temporal Filtering (MCTF) \cite{garbasTCSVT}.
Blur and ghosting artifacts caused by displacement in the signal can
be reduced and thus the visual quality of the lowpass band is increased.
A high quality of the lowpass band is very important when it shall
be used as a scalable representation.

Combined with a feasible wavelet coefficient coder, a scalable representation
of parts of the volume can be obtained \cite{sanchez2010}. Especially
with anatomic segmentation information available \cite{cavallaro2011region},
applications are possible, where different parts of a volume, like
the heart or the lung, can be decoded with a higher quality while
the remaining volume can be decoded in a coarser representation. Thus,
only the important information has to be decoded. Integer implementation
of the wavelet transform allows a scalable representation where lossless
reconstruction of the original volume is possible.

Like in wavelet-based video coding, motion compensation methods can
be used for a compensated wavelet transform of volumes from Computed
Tomography \cite{schnurrer2012vcip}.  In \cite{schnurrer2012mmsp}
we show that a \zweid{} \meshbased{} compensation method is more
feasible to model the deforming displacement over time than a \blockbased{}
method that is usually used in video coding. The first row of \Fig{\ref{fig:Lifting-structure}}
illustrates a \dreidt{} volume, that means a \dreid{} volume changing
over time. To obtain a predictor for a volume shown on the left side,
a \zweid{} motion compensation method (MC) can be applied subsequently
for all slices of a time instant. The disadvantage of a \zweid{}
method is that a displacement in slice-direction cannot be compensated.
To overcome this limitation, we propose a \dreid{} method as shown
on the right side.

Section~2 presents a brief review of the compensated wavelet lifting
followed by a detailed description of the mesh compensation in Section~3.
The estimation of the grid point motion is a crucial task. We introduce
a modification to the motion estimation for the \zweid{} mesh to
avoid the degeneration of the mesh structure and thus obtain an improved
inversion. Then, the extension to \dreid{} is presented. Simulation
results are presented in Section~4.

\section{Compensated Wavelet Lifting}

\begin{figure}
\begin{centering}

\psfragscanon

\psfrag{time}{time}
\psfrag{twod}{\zweid}
\psfrag{thrd}{\dreid}
\psfrag{MC}{MC}
\psfrag{IMC}{IMC}
\psfrag{hp}{$\HP_{t}$}
\psfrag{lp}{$\LP_{t}$}
\psfrag{time}{time $t$}
\psfrag{x}{$x$}
\psfrag{y}{$y$}
\psfrag{z}{$z$}
\psfrag{m}{$-$}
\psfrag{p}{$\frac{1}{2}$}

\psfrag{f1}{$f_{2t\hspace{-0.2ex}-\hspace{-0.2ex}1}$}

\psfrag{f2}{$f_{2t}$}

\includegraphics[width=0.99\columnwidth]{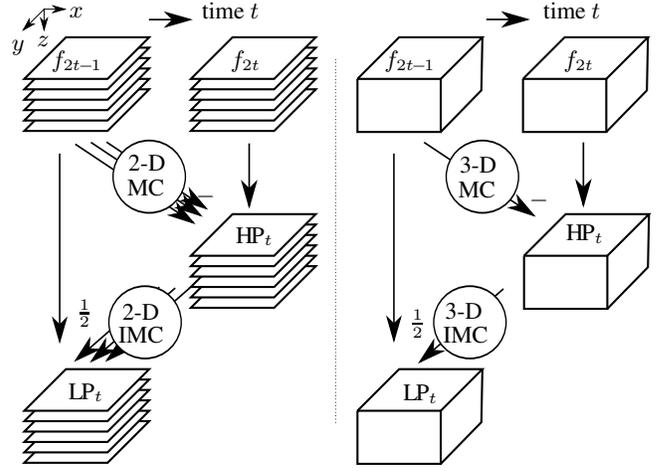}

\psfragscanoff{}

\end{centering}

\spaceBeforeLabel\caption{\label{fig:Lifting-structure}Compensated Haar lifting structure in
temporal direction (MCTF) with \zweid{} compensation (left) and \dreid{}
compensation (right)}

\spaceBelowFig
\end{figure}

\Fig{\ref{fig:Lifting-structure}} shows the lifting structure of
the Haar wavelet transform extended by a compensation method. The
transform is applied in temporal direction. The volumes $\sigf_{t}$
are indexed by the time step $t$ in temporal direction. The lifting
structure consists of a prediction step and an update step. The highpass
coefficients $\HP_{t}$ are computed in 
\begin{figure*}[!t]
\psfragscanon

\psfrag{A}{$A$}

\psfrag{P}{$P$}

\psfrag{point}{$P$}

\psfrag{B}{$B$}

\psfrag{C}{$C$}

\psfrag{D}{$D$}

\psfrag{dist}{$d$}

\includegraphics[width=0.24\textwidth]{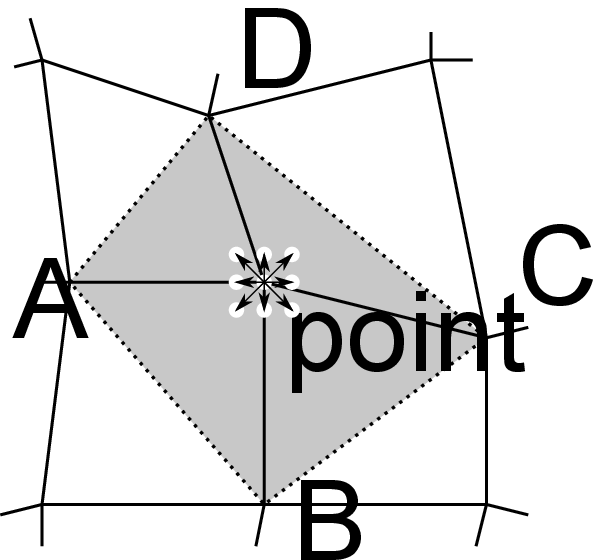}\hfill{}\includegraphics[width=0.24\textwidth]{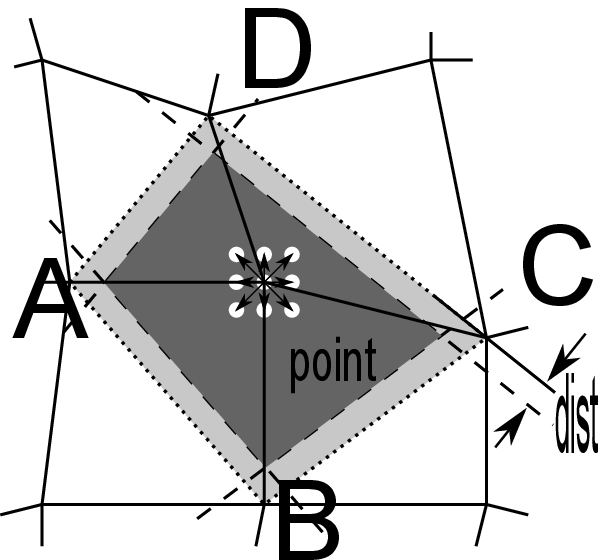}\hfill{}\includegraphics[width=0.24\textwidth]{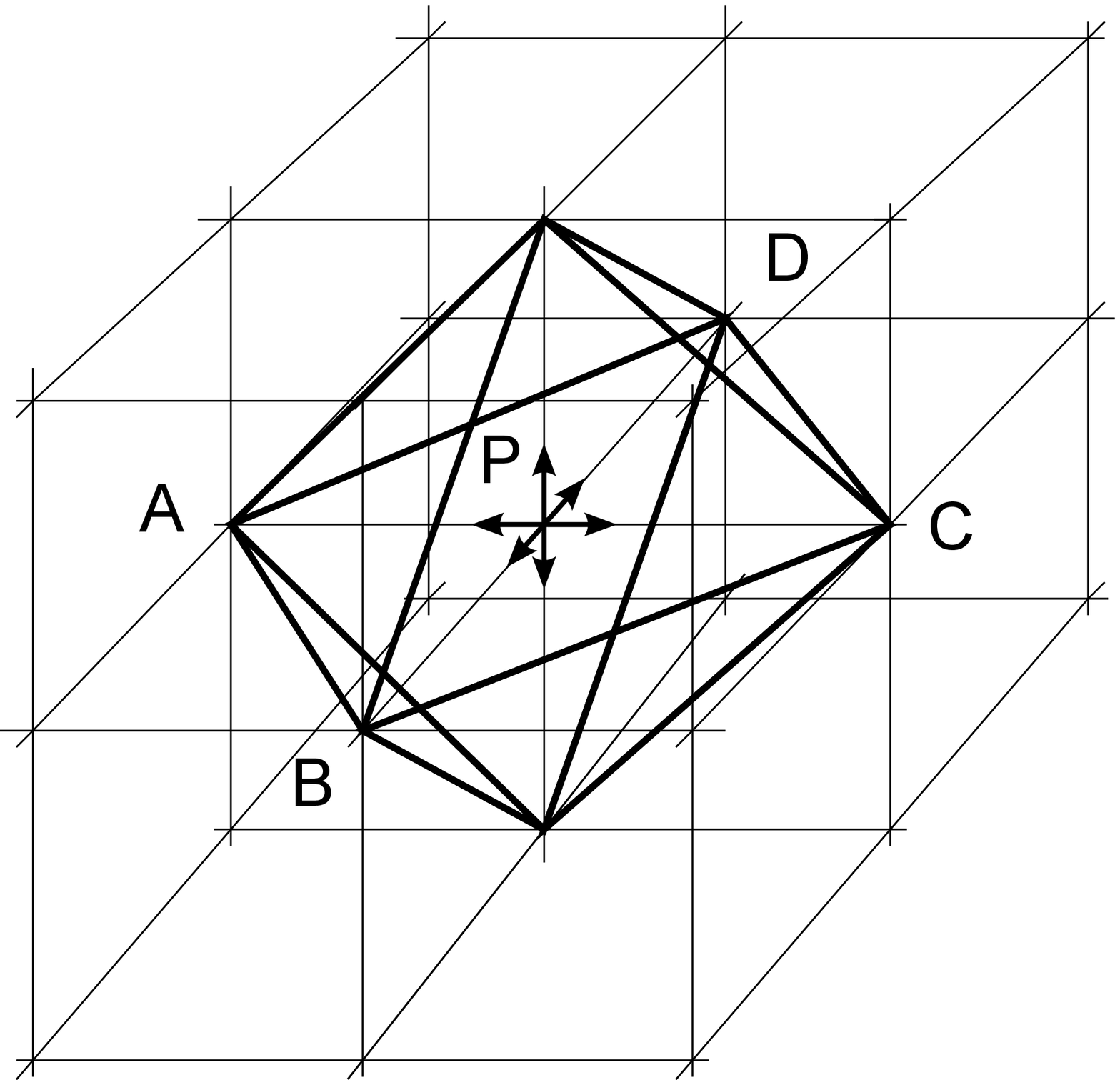}\hfill{}\includegraphics[width=0.24\textwidth]{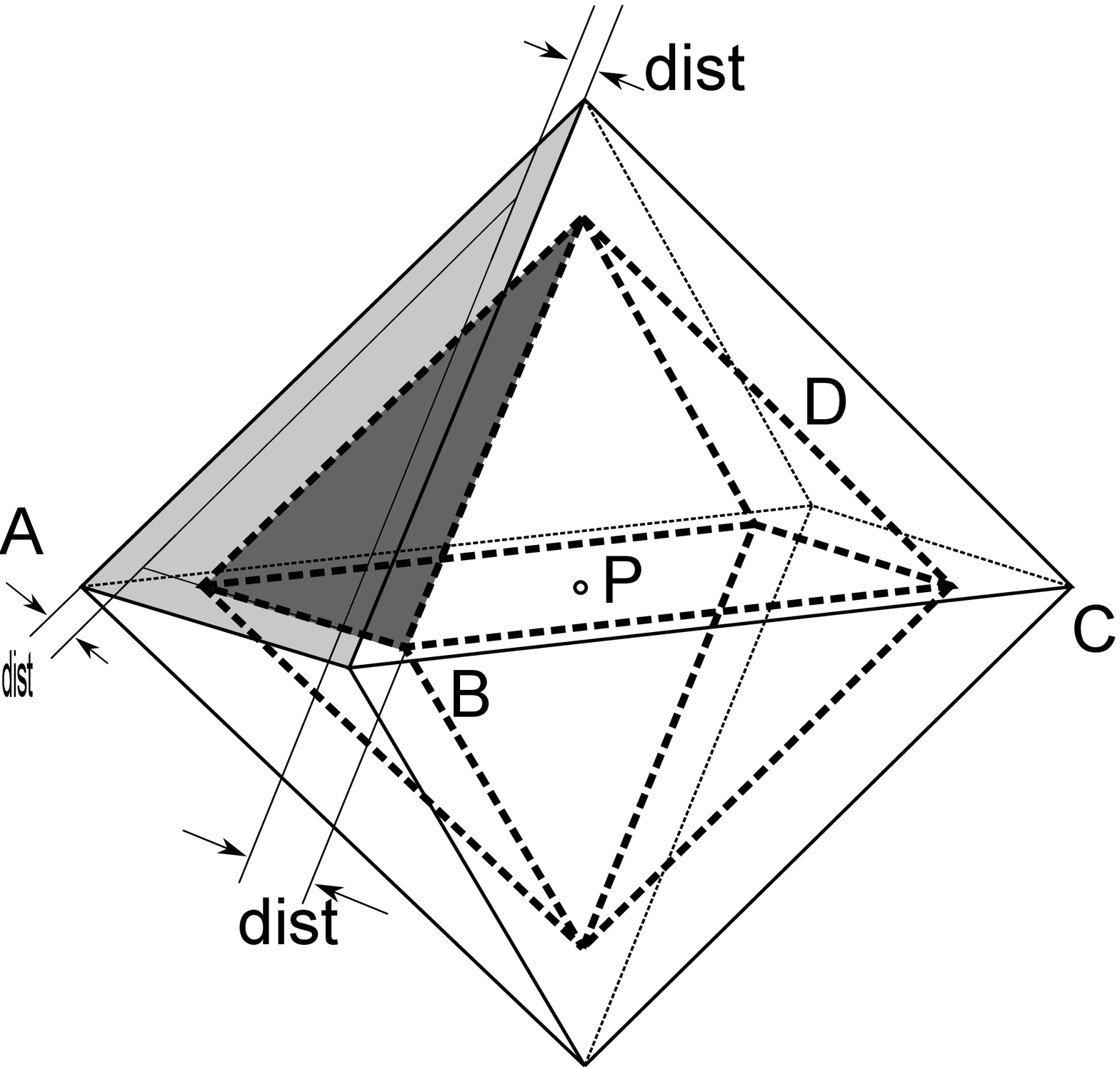}

\spaceBeforeLabel{}\hfill{}(a)\hfill{}\hfill{}(b)\hfill{}\hfill{}(c)\hfill{}\hfill{}(d)\hfill{}

\psfragscanoff

\spaceBeforeLabel{}\spaceBeforeLabel{}\spaceBeforeLabel{}

\caption{\label{fig:independentGridPoints}Illustration of the refinement search
range for the grid points of the \zweid{} mesh (a) extended by an
additional safety boundary (b) and the refinement search range for
the \dreid{} mesh (c). Search area of the \dreid{} mesh with additional
safety boundary $d$, illustrated for one limiting plane in gray (d),
resulting in the dashed octahedron}

\spaceBelowFig
\end{figure*}
 the prediction step according to
\begin{equation}
\HP_{t}=\sigf_{2t}-\left\lfloor \mathcal{W}_{2t-1\rightarrow2t}\left(\sigf_{2t-1}\right)\right\rfloor .\label{eq:H-Haar}
\end{equation}
Instead of the original volume $\sigf_{2t-1}$, a predictor is subtracted
from $\sigf_{2t}$, denoted by the warping operator $\mathcal{W}_{2t-1\rightarrow2t}$
\cite{garbasTCSVT}. The compensation is called MC in \Fig{\ref{fig:Lifting-structure}}.
In the update step, the lowpass coefficients $\LP_{t}$ are computed
by 
\begin{equation}
\LP_{t}=\sigf_{2t-1}+\left\lfloor \frac{1}{2}\mathcal{W}_{2t\rightarrow2t-1}\left(\HP_{t}\right)\right\rfloor .\label{eq:L-Haar}
\end{equation}
As the index of $\mathcal{W}$$ $ in \eqref{eq:L-Haar} shows, the
compensation has to be inverted in the update step to achieve an equivalent
wavelet transform. The inversion is denoted by IMC in \Fig{\ref{fig:Lifting-structure}}.
For a compensated wavelet transform, the \meshbased{} compensation
method has the advantage that it is invertible \cite{secker2002}.
 To avoid rounding errors, rounding operations are applied to the
fractional parts \cite{schnurrer2012mmsp,calderbank1997}. The reconstruction
of the original volume without any loss is very important, e.g., for
medical image data.

\section{Mesh-based Compensation}

In \zweid{}, a mesh-based compensation is computed by putting a mesh
over the reference image. The predictor is then computed by warping
the reference image according to the deformation given by the motion
vectors of the grid points \cite{schnurrer2012mmsp}. A quadrilateral
mesh topology leads to a bilinear transform \cite{nakaya1994} of
the patches of the underlying image.  Compared to an affine transform
of a triangle mesh, the quadrilateral mesh leads to a smoother motion
vector field \cite{schnurrer2012mmsp} and thus a smoother prediction.
The displacement in the \dreidt{} volumes can be well modeled by
smooth representations \cite{Weinlich2012}.

\subsection{Motion Estimation in \zweid{}}

The estimation of the grid point (GP) motion is a crucial task. \Fig{\ref{fig:independentGridPoints}}~(a)
shows a detail of a quadrilateral mesh. The motion vectors between
the GPs are computed from an interpolation. Thus, the choice of the
motion vector of $P$ in the center is influenced by its eight surrounding
neighbors. A modification of $P$ influences all surrounding neighbors
as well. An optimum solution is a combination of all motion vectors
of all GPs which is computationally too complex. Several iterative
methods exist and we use an iterative refinement process as proposed
in \cite{nakaya1994,sullivan1991}. We initialize the motion vectors
with zeros. In every iteration, the current motion vector of each
GP is updated in every direction, the corresponding warping of the
underlying image is computed and the update with the smallest error
metric is chosen. The nine update positions for $P$ are illustrated
in \Fig{\ref{fig:independentGridPoints}~(a)}. Independent GPs can
be refined in parallel \cite{nakaya1994}. In order to avoid a degeneration
of the mesh structure, the movement of the GPs can be limited to a
specific search range \cite{wang2012}. This search range is limited
by the direct neighbors, labeled by $A$, $B$, $C$, and $D$. It
is marked light gray in \Fig{\ref{fig:independentGridPoints}~(a)}.
This prevents concave quadrilateral structures in the mesh that make
a proper inversion very complicated or even impossible \cite{wang1996p1}.
We introduce a further safety boundary $d$ to prevent the degradation
to triangular structures. The resulting search area is illustrated
in \Fig{\ref{fig:independentGridPoints}}~(b) in dark gray. Movements
of $P$ have to stay inside the dark gray area. With this extension,
a convex mesh structure is maintained. This is advantageous for the
inversion of the compensation in the update step.

To avoid the occurrence of unconnected pixels \cite{ohm2004interframe,girod2005,schnurrer2013},
the movement of the boundary GPs of the complete mesh is further limited.
The shape of the outer hull shall remain, so the four corner GPs are
kept fix, i.e., no motion is allowed. The remaining GPs on the left
and right boundary edge can only move up and down while the GPs on
the top and bottom boundary edge can only move left and right. The
\zweid{} mesh corresponds to a subsampling of a \zweid{} motion
vector field where the motion vectors of the intermediate positions
are obtained by bilinear interpolation.

\subsection{Proposed Extension to \dreid{}}

The mesh is extended by one dimension to obtain a \dreid{} compensation
method. This corresponds to a further subsampling of the motion vector
field in the third dimension. For the \dreid{} mesh, a motion vector
of a GP has three components. The motion vectors of the intermediate
positions are obtained by trilinear interpolation. A detail of a \dreid{}
mesh is illustrated in \Fig{\ref{fig:independentGridPoints}~(c)}.
To estimate the motion vectors of the GPs, a similar method as for
the \zweid{} mesh is applied. The refinement for the current GP is
tested in the third dimension as well, resulting in 27 test positions.
In this way, deformation in the third dimension is compensated. For
every update position, the warping of the corresponding image cube
is computed. The  update for the motion vector with the smallest
error metric is chosen.

Again, to avoid a degeneration of the mesh structure, the update positions
are limited to a specific search area. In the \dreid{} case, this
search area results in an octahedron marked by bold black lines in
\Fig{\ref{fig:independentGridPoints}~(c)}. For illustration, the
octahedron of the search area in \Fig{\ref{fig:independentGridPoints}~(c)}
is again shown in \Fig{\ref{fig:independentGridPoints}~(d)} in a
different perspective. We add a safety boundary $d$ for the \dreid{}
case as well. This is shown exemplarily for one plane of the octahedron.
$P$ has to remain in the smaller dashed octahedron to maintain a
proper mesh structure.

The movement of the GPs on the boundary is limited similar to the
\zweid{} case to maintain the shape of the outer hull of the mesh.
For instance, the GPs on the front and the back boundary face can
only move up, down, left, and right.

\begin{figure}[!t]
% Autor: Peter Scholz
% Email: contact@peter-scholz.net
% Date:  27-Jan-2014 22:11:18
%
% This file was created by fig2texPS. Note, that the packages
% pstricks, pst-node, pst-plot, pst-circ and moredefs are required.
% A minimal example code could be:
%
% \documentclass{article}
% \usepackage{pstricks, pst-node, pst-plot, pst-circ}
% \usepackage{moredefs}
% \begin{document}
% \input{fig1.tex}
% \end{document}
%
% Global Parameters that can be changed:
\providelength{\AxesLineWidth}       \setlength{\AxesLineWidth}{0.5pt}%
\providelength{\plotwidth}           \setlength{\plotwidth}{7cm}% width of the axes only
\providelength{\LineWidth}           \setlength{\LineWidth}{0.7pt}%
\providelength{\MarkerSize}          \setlength{\MarkerSize}{4pt}%
\newrgbcolor{GridColor}{0.8 0.8 0.8}%
%
% Begin Figure:-------------------------------------------
\psset{xunit=0.022222\plotwidth,yunit=0.000225\plotwidth}%
\begin{pspicture}(-2.392857,556.237219)(50.064286,4506.339468)%

% Draw bounding box for test aspects: ----
% \psframe(-2.392857,556.237219)(50.064286,4506.339468)
% Total width:  8.160000 cm
% Total height: 6.230968 cm

% Draw Ticks: ----
% x-Ticks:
\psline[linewidth=\AxesLineWidth,linecolor=GridColor](5.000000,1000.000000)(5.000000,1053.251534)
\psline[linewidth=\AxesLineWidth,linecolor=GridColor](10.000000,1000.000000)(10.000000,1053.251534)
\psline[linewidth=\AxesLineWidth,linecolor=GridColor](15.000000,1000.000000)(15.000000,1053.251534)
\psline[linewidth=\AxesLineWidth,linecolor=GridColor](20.000000,1000.000000)(20.000000,1053.251534)
\psline[linewidth=\AxesLineWidth,linecolor=GridColor](25.000000,1000.000000)(25.000000,1053.251534)
\psline[linewidth=\AxesLineWidth,linecolor=GridColor](30.000000,1000.000000)(30.000000,1053.251534)
\psline[linewidth=\AxesLineWidth,linecolor=GridColor](35.000000,1000.000000)(35.000000,1053.251534)
\psline[linewidth=\AxesLineWidth,linecolor=GridColor](40.000000,1000.000000)(40.000000,1053.251534)
\psline[linewidth=\AxesLineWidth,linecolor=GridColor](45.000000,1000.000000)(45.000000,1053.251534)
\psline[linewidth=\AxesLineWidth,linecolor=GridColor](50.000000,1000.000000)(50.000000,1053.251534)
% y-Ticks:
\psline[linewidth=\AxesLineWidth,linecolor=GridColor](5.000000,1000.000000)(5.540000,1000.000000)
\psline[linewidth=\AxesLineWidth,linecolor=GridColor](5.000000,1500.000000)(5.540000,1500.000000)
\psline[linewidth=\AxesLineWidth,linecolor=GridColor](5.000000,2000.000000)(5.540000,2000.000000)
\psline[linewidth=\AxesLineWidth,linecolor=GridColor](5.000000,2500.000000)(5.540000,2500.000000)
\psline[linewidth=\AxesLineWidth,linecolor=GridColor](5.000000,3000.000000)(5.540000,3000.000000)
\psline[linewidth=\AxesLineWidth,linecolor=GridColor](5.000000,3500.000000)(5.540000,3500.000000)
\psline[linewidth=\AxesLineWidth,linecolor=GridColor](5.000000,4000.000000)(5.540000,4000.000000)
\psline[linewidth=\AxesLineWidth,linecolor=GridColor](5.000000,4500.000000)(5.540000,4500.000000)

{ \footnotesize % FontSizeTickLabels
% Draw x-Labels: ----
\rput[t](5.000000,946.748466){$5$}
\rput[t](10.000000,946.748466){$10$}
\rput[t](15.000000,946.748466){$15$}
\rput[t](20.000000,946.748466){$20$}
\rput[t](25.000000,946.748466){$25$}
\rput[t](30.000000,946.748466){$30$}
\rput[t](35.000000,946.748466){$35$}
\rput[t](40.000000,946.748466){$40$}
\rput[t](45.000000,946.748466){$45$}
\rput[t](50.000000,946.748466){$50$}
% Draw y-Labels: ----
\rput[r](4.460000,1000.000000){$1000$}
\rput[r](4.460000,1500.000000){$1500$}
\rput[r](4.460000,2000.000000){$2000$}
\rput[r](4.460000,2500.000000){$2500$}
\rput[r](4.460000,3000.000000){$3000$}
\rput[r](4.460000,3500.000000){$3500$}
\rput[r](4.460000,4000.000000){$4000$}
\rput[r](4.460000,4500.000000){$4500$}
} % End FontSizeTickLabels

% Draw Axes: ----
\psline[linewidth=\AxesLineWidth](5.000000,1000.000000)(50.000000,1000.000000)
\psline[linewidth=\AxesLineWidth](5.000000,1000.000000)(5.000000,4500.000000)

{ \small % FontSizeXYlabel
% x-Label: ----
\rput[b](27.500000,556.237219){
\begin{tabular}{c}
Iteration\\
\end{tabular}
}

% y-Label: ----
\rput[t]{90}(-2.392857,2750.000000){
\begin{tabular}{c}
\MSE{} $\HP_1$\\
\end{tabular}
}
} % End FontSizeXYlabel

% New Line DATA: ----
\newrgbcolor{color481.0016}{0  0  1}
\psline[plotstyle=line,linejoin=1,linestyle=dashed,dash=2pt 3pt,linewidth=\LineWidth,linecolor=color481.0016]
(5.000000,4231.486043)(10.000000,4231.486043)(12.000000,4231.486043)(14.000000,4231.486043)(16.000000,4231.486043)
(18.000000,4231.486043)(20.000000,4231.486043)(22.000000,4231.486043)(24.000000,4231.486043)(30.000000,4231.486043)
(35.000000,4231.486043)(40.000000,4231.486043)(45.000000,4231.486043)(50.000000,4231.486043)

% New Line DATA: ----
\newrgbcolor{color482.0011}{1  0  0}
\psline[plotstyle=line,linejoin=1,showpoints=true,dotstyle=Bsquare,dotsize=\MarkerSize,linestyle=solid,linewidth=\LineWidth,linecolor=color482.0011]
(5.000000,2235.529020)(10.000000,2180.840355)(12.000000,2175.990820)(14.000000,2173.945011)(16.000000,2171.990755)
(18.000000,2170.985948)(20.000000,2169.798302)(22.000000,2169.024002)(24.000000,2167.514583)(30.000000,2167.557843)
(35.000000,2166.886278)(40.000000,2166.821207)(45.000000,2166.126087)(50.000000,2165.652207)

% New Line DATA: ----
\newrgbcolor{color483.0011}{0  1  1}
\psline[plotstyle=line,linejoin=1,showpoints=true,dotstyle=Bo,dotsize=\MarkerSize,linestyle=solid,linewidth=\LineWidth,linecolor=color483.0011]
(5.000000,1816.027683)(10.000000,1764.671960)(12.000000,1760.082650)(14.000000,1757.001019)(16.000000,1754.870447)
(18.000000,1753.431103)(20.000000,1752.641103)(22.000000,1751.880914)(24.000000,1751.113839)(30.000000,1749.407485)
(35.000000,1748.579414)(40.000000,1747.208763)(45.000000,1746.182620)(50.000000,1745.925408)

% New Line DATA: ----
\newrgbcolor{color484.0011}{1  0  1}
\psline[plotstyle=line,linejoin=1,showpoints=true,dotstyle=Btriangle,dotsize=\MarkerSize,linestyle=solid,linewidth=\LineWidth,linecolor=color484.0011]
(5.000000,1554.190374)(10.000000,1499.962387)(12.000000,1494.778489)(14.000000,1491.902942)(16.000000,1489.543497)
(18.000000,1487.766956)(20.000000,1486.017872)(22.000000,1484.771467)(24.000000,1483.752572)(30.000000,1482.220791)
(35.000000,1482.049141)(40.000000,1481.562766)(45.000000,1481.205518)(50.000000,1481.099757)

% New Line DATA: ----
\newrgbcolor{color485.0011}{0  1  0}
\psline[plotstyle=line,linejoin=1,linestyle=dashed,linewidth=\LineWidth,linecolor=color485.0011]
(5.000000,1772.999485)(10.000000,1614.650819)(12.000000,1592.236290)(14.000000,1578.063171)(16.000000,1568.113208)
(18.000000,1561.623073)(20.000000,1557.021257)(22.000000,1553.606410)(24.000000,1551.296827)(30.000000,1547.208816)
(35.000000,1545.450490)(40.000000,1544.433093)(45.000000,1543.719517)(50.000000,1543.400584)

% Legend: ----
{ \small % FontSizeLegend
\rput[r](48.920000,2950.000000){%
\psframebox[framesep=0pt,linewidth=\AxesLineWidth]{\psframebox*{\begin{tabular}{l}
\Rnode{a1}{\hspace*{0.0ex}} \hspace*{0.7cm} \Rnode{a2}{~~no compensation} \\
\Rnode{a3}{\hspace*{0.0ex}} \hspace*{0.7cm} \Rnode{a4}{~~3-D mesh, $16\times16\times16$} \\
\Rnode{a5}{\hspace*{0.0ex}} \hspace*{0.7cm} \Rnode{a6}{~~3-D mesh, $16\times16\times8$} \\
\Rnode{a7}{\hspace*{0.0ex}} \hspace*{0.7cm} \Rnode{a8}{~~3-D mesh, $16\times16\times4$} \\
\Rnode{a9}{\hspace*{0.0ex}} \hspace*{0.7cm} \Rnode{a10}{~~2-D mesh, $16\times16$} \\
\end{tabular}}
\ncline[linestyle=dashed,dash=2pt 3pt,linewidth=\LineWidth,linecolor=color481.0016]{a1}{a2}
\ncline[linestyle=solid,linewidth=\LineWidth,linecolor=color482.0011]{a3}{a4} \ncput{\psdot[dotstyle=Bsquare,dotsize=\MarkerSize,linecolor=color482.0011]}
\ncline[linestyle=solid,linewidth=\LineWidth,linecolor=color483.0011]{a5}{a6} \ncput{\psdot[dotstyle=Bo,dotsize=\MarkerSize,linecolor=color483.0011]}
\ncline[linestyle=solid,linewidth=\LineWidth,linecolor=color484.0011]{a7}{a8} \ncput{\psdot[dotstyle=Btriangle,dotsize=\MarkerSize,linecolor=color484.0011]}
\ncline[linestyle=dashed,linewidth=\LineWidth,linecolor=color485.0011]{a9}{a10}
}%
}%
} % End FontSizeLegend

\end{pspicture}%

\spaceBeforeLabel\spaceBeforeLabel\spaceBeforeLabel

\caption{\label{fig:hpmse}The mean squared energy of the highpass band $\HP_{1}$
for no compensation, \zweid{} mesh and \dreid{} mesh is plotted
against the refinement iteration used for the \meshbased{} compensation
methods}

\spaceBelowFig
\end{figure}
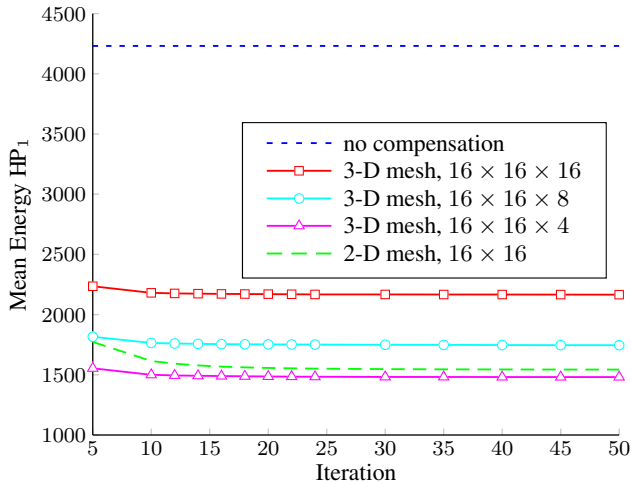

\subsection{Inversion for the Update Step}

In the update step of the compensated lifting \eqref{eq:L-Haar},
the compensation from the prediction step \eqref{eq:H-Haar} has to
be inverted to obtain an equivalent wavelet transform \cite{schnurrer2012mmsp}.
The \meshbased{} approach has the advantage that it is invertible
\cite{secker2002}. We use  an approximation from \cite{secker2002}
for the inversion of the \meshbased{} compensation, thus accepting
the induced error. Instead of calculating the inversion of the mesh
warping, we take the negative values of the motion vectors at the
GPs.

\section{Simulation Results}

\label{sec:Simulation}For evaluating our compensation methods, we
used a \emph{cardiac} \dreidt{} CT data set%
\footnote{The CT volume data set was kindly provided by Siemens Healthcare.%
}. This multidimensional \dreidt{} volume has 10 time steps, 128 slices
in $z$-direction and a resolution of $512\times512$ pixels in $xy$-direction
at 12~bit per sample and shows a beating heart over time.

The safety boundary is chosen to $d=1$ pixel. For the \zweid{} mesh,
we used a grid size of $16\times16$ pixels. This size has proven
to be reasonable in \cite{schnurrer2012mmsp}. This corresponds to
a subsampling of the motion vector field by $1:16^{2}=1:256$ in $xy$-direction.
This grid size is used for the \dreid{} mesh as well in $xy$-direction.
For the grid size in $z$-direction, 16, 8 and 4 pixels were used
to test different subsampling factors in $z$-direction.

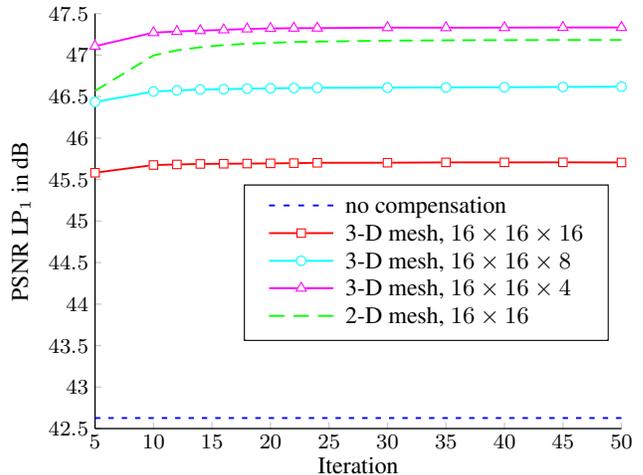
\begin{figure}[!t]
% Autor: Peter Scholz
% Email: contact@peter-scholz.net
% Date:  27-Jan-2014 22:11:16
%
% This file was created by fig2texPS. Note, that the packages
% pstricks, pst-node, pst-plot, pst-circ and moredefs are required.
% A minimal example code could be:
%
% \documentclass{article}
% \usepackage{pstricks, pst-node, pst-plot, pst-circ}
% \usepackage{moredefs}
% \begin{document}
% \input{fig1.tex}
% \end{document}
%
% Global Parameters that can be changed:
\providelength{\AxesLineWidth}       \setlength{\AxesLineWidth}{0.5pt}%
\providelength{\plotwidth}           \setlength{\plotwidth}{7cm}% width of the axes only
\providelength{\LineWidth}           \setlength{\LineWidth}{0.7pt}%
\providelength{\MarkerSize}          \setlength{\MarkerSize}{4pt}%
\newrgbcolor{GridColor}{0.8 0.8 0.8}%
%
% Begin Figure:-------------------------------------------
\psset{xunit=0.022222\plotwidth,yunit=0.157742\plotwidth}%
\begin{pspicture}(-2.392857,41.866053)(50.064286,47.509056)%

% Draw bounding box for test aspects: ----
% \psframe(-2.392857,41.866053)(50.064286,47.509056)
% Total width:  8.160000 cm
% Total height: 6.230968 cm

% Draw Ticks: ----
% x-Ticks:
\psline[linewidth=\AxesLineWidth,linecolor=GridColor](5.000000,42.500000)(5.000000,42.576074)
\psline[linewidth=\AxesLineWidth,linecolor=GridColor](10.000000,42.500000)(10.000000,42.576074)
\psline[linewidth=\AxesLineWidth,linecolor=GridColor](15.000000,42.500000)(15.000000,42.576074)
\psline[linewidth=\AxesLineWidth,linecolor=GridColor](20.000000,42.500000)(20.000000,42.576074)
\psline[linewidth=\AxesLineWidth,linecolor=GridColor](25.000000,42.500000)(25.000000,42.576074)
\psline[linewidth=\AxesLineWidth,linecolor=GridColor](30.000000,42.500000)(30.000000,42.576074)
\psline[linewidth=\AxesLineWidth,linecolor=GridColor](35.000000,42.500000)(35.000000,42.576074)
\psline[linewidth=\AxesLineWidth,linecolor=GridColor](40.000000,42.500000)(40.000000,42.576074)
\psline[linewidth=\AxesLineWidth,linecolor=GridColor](45.000000,42.500000)(45.000000,42.576074)
\psline[linewidth=\AxesLineWidth,linecolor=GridColor](50.000000,42.500000)(50.000000,42.576074)
% y-Ticks:
\psline[linewidth=\AxesLineWidth,linecolor=GridColor](5.000000,42.500000)(5.540000,42.500000)
\psline[linewidth=\AxesLineWidth,linecolor=GridColor](5.000000,43.000000)(5.540000,43.000000)
\psline[linewidth=\AxesLineWidth,linecolor=GridColor](5.000000,43.500000)(5.540000,43.500000)
\psline[linewidth=\AxesLineWidth,linecolor=GridColor](5.000000,44.000000)(5.540000,44.000000)
\psline[linewidth=\AxesLineWidth,linecolor=GridColor](5.000000,44.500000)(5.540000,44.500000)
\psline[linewidth=\AxesLineWidth,linecolor=GridColor](5.000000,45.000000)(5.540000,45.000000)
\psline[linewidth=\AxesLineWidth,linecolor=GridColor](5.000000,45.500000)(5.540000,45.500000)
\psline[linewidth=\AxesLineWidth,linecolor=GridColor](5.000000,46.000000)(5.540000,46.000000)
\psline[linewidth=\AxesLineWidth,linecolor=GridColor](5.000000,46.500000)(5.540000,46.500000)
\psline[linewidth=\AxesLineWidth,linecolor=GridColor](5.000000,47.000000)(5.540000,47.000000)
\psline[linewidth=\AxesLineWidth,linecolor=GridColor](5.000000,47.500000)(5.540000,47.500000)

{ \footnotesize % FontSizeTickLabels
% Draw x-Labels: ----
\rput[t](5.000000,42.423926){$5$}
\rput[t](10.000000,42.423926){$10$}
\rput[t](15.000000,42.423926){$15$}
\rput[t](20.000000,42.423926){$20$}
\rput[t](25.000000,42.423926){$25$}
\rput[t](30.000000,42.423926){$30$}
\rput[t](35.000000,42.423926){$35$}
\rput[t](40.000000,42.423926){$40$}
\rput[t](45.000000,42.423926){$45$}
\rput[t](50.000000,42.423926){$50$}
% Draw y-Labels: ----
\rput[r](4.460000,42.500000){$42.5$}
\rput[r](4.460000,43.000000){$43$}
\rput[r](4.460000,43.500000){$43.5$}
\rput[r](4.460000,44.000000){$44$}
\rput[r](4.460000,44.500000){$44.5$}
\rput[r](4.460000,45.000000){$45$}
\rput[r](4.460000,45.500000){$45.5$}
\rput[r](4.460000,46.000000){$46$}
\rput[r](4.460000,46.500000){$46.5$}
\rput[r](4.460000,47.000000){$47$}
\rput[r](4.460000,47.500000){$47.5$}
} % End FontSizeTickLabels

% Draw Axes: ----
\psline[linewidth=\AxesLineWidth](5.000000,42.500000)(50.000000,42.500000)
\psline[linewidth=\AxesLineWidth](5.000000,42.500000)(5.000000,47.500000)

{ \small % FontSizeXYlabel
% x-Label: ----
\rput[b](27.500000,41.866053){
\begin{tabular}{c}
Iteration\\
\end{tabular}
}

% y-Label: ----
\rput[t]{90}(-2.392857,45.000000){
\begin{tabular}{c}
PSNR $\LP_1$ in dB\\
\end{tabular}
}
} % End FontSizeXYlabel

% New Line DATA: ----
\newrgbcolor{color241.0016}{0  0  1}
\psline[plotstyle=line,linejoin=1,linestyle=dashed,dash=2pt 3pt,linewidth=\LineWidth,linecolor=color241.0016]
(5.000000,42.626893)(10.000000,42.626893)(12.000000,42.626893)(14.000000,42.626893)(16.000000,42.626893)
(18.000000,42.626893)(20.000000,42.626893)(22.000000,42.626893)(24.000000,42.626893)(30.000000,42.626893)
(35.000000,42.626893)(40.000000,42.626893)(45.000000,42.626893)(50.000000,42.626893)

% New Line DATA: ----
\newrgbcolor{color242.0011}{1  0  0}
\psline[plotstyle=line,linejoin=1,showpoints=true,dotstyle=Bsquare,dotsize=\MarkerSize,linestyle=solid,linewidth=\LineWidth,linecolor=color242.0011]
(5.000000,45.582473)(10.000000,45.675249)(12.000000,45.682008)(14.000000,45.688277)(16.000000,45.690526)
(18.000000,45.692579)(20.000000,45.695193)(22.000000,45.697667)(24.000000,45.702275)(30.000000,45.703033)
(35.000000,45.707044)(40.000000,45.707091)(45.000000,45.707772)(50.000000,45.706140)

% New Line DATA: ----
\newrgbcolor{color243.0011}{0  1  1}
\psline[plotstyle=line,linejoin=1,showpoints=true,dotstyle=Bo,dotsize=\MarkerSize,linestyle=solid,linewidth=\LineWidth,linecolor=color243.0011]
(5.000000,46.434904)(10.000000,46.561309)(12.000000,46.574117)(14.000000,46.584911)(16.000000,46.588782)
(18.000000,46.594858)(20.000000,46.599154)(22.000000,46.602934)(24.000000,46.605187)(30.000000,46.607795)
(35.000000,46.610330)(40.000000,46.612949)(45.000000,46.616033)(50.000000,46.619670)

% New Line DATA: ----
\newrgbcolor{color244.0011}{1  0  1}
\psline[plotstyle=line,linejoin=1,showpoints=true,dotstyle=Btriangle,dotsize=\MarkerSize,linestyle=solid,linewidth=\LineWidth,linecolor=color244.0011]
(5.000000,47.107349)(10.000000,47.270269)(12.000000,47.284522)(14.000000,47.294243)(16.000000,47.304275)
(18.000000,47.314918)(20.000000,47.320688)(22.000000,47.324201)(24.000000,47.325241)(30.000000,47.330875)
(35.000000,47.329692)(40.000000,47.331006)(45.000000,47.332687)(50.000000,47.332324)

% New Line DATA: ----
\newrgbcolor{color245.0011}{0  1  0}
\psline[plotstyle=line,linejoin=1,linestyle=dashed,linewidth=\LineWidth,linecolor=color245.0011]
(5.000000,46.568787)(10.000000,46.995617)(12.000000,47.057373)(14.000000,47.094318)(16.000000,47.120566)
(18.000000,47.136303)(20.000000,47.148301)(22.000000,47.157541)(24.000000,47.162709)(30.000000,47.173866)
(35.000000,47.178056)(40.000000,47.180208)(45.000000,47.181675)(50.000000,47.182094)

% Legend: ----
{ \small % FontSizeLegend
\rput[r](48.920000,44.500000){%
\psframebox[framesep=0pt,linewidth=\AxesLineWidth]{\psframebox*{\begin{tabular}{l}
\Rnode{a1}{\hspace*{0.0ex}} \hspace*{0.7cm} \Rnode{a2}{~~no compensation} \\
\Rnode{a3}{\hspace*{0.0ex}} \hspace*{0.7cm} \Rnode{a4}{~~3-D mesh, $16\times16\times16$} \\
\Rnode{a5}{\hspace*{0.0ex}} \hspace*{0.7cm} \Rnode{a6}{~~3-D mesh, $16\times16\times8$} \\
\Rnode{a7}{\hspace*{0.0ex}} \hspace*{0.7cm} \Rnode{a8}{~~3-D mesh, $16\times16\times4$} \\
\Rnode{a9}{\hspace*{0.0ex}} \hspace*{0.7cm} \Rnode{a10}{~~2-D mesh, $16\times16$} \\
\end{tabular}}
\ncline[linestyle=dashed,dash=2pt 3pt,linewidth=\LineWidth,linecolor=color241.0016]{a1}{a2}
\ncline[linestyle=solid,linewidth=\LineWidth,linecolor=color242.0011]{a3}{a4} \ncput{\psdot[dotstyle=Bsquare,dotsize=\MarkerSize,linecolor=color242.0011]}
\ncline[linestyle=solid,linewidth=\LineWidth,linecolor=color243.0011]{a5}{a6} \ncput{\psdot[dotstyle=Bo,dotsize=\MarkerSize,linecolor=color243.0011]}
\ncline[linestyle=solid,linewidth=\LineWidth,linecolor=color244.0011]{a7}{a8} \ncput{\psdot[dotstyle=Btriangle,dotsize=\MarkerSize,linecolor=color244.0011]}
\ncline[linestyle=dashed,linewidth=\LineWidth,linecolor=color245.0011]{a9}{a10}
}%
}%
} % End FontSizeLegend

\end{pspicture}%

\spaceBeforeLabel\spaceBeforeLabel\spaceBeforeLabel

\caption{\label{fig:lppsnr}The quality of the lowpass band in terms of $\text{PSNR}\left(\LP_{1},f_{1}\right)$
in dB for no compensation, \zweid{} mesh and \dreid{} mesh is plotted
against the refinement iterations used for the \meshbased{} compensation
methods}

\spaceBelowFig
\end{figure}

\Fig{\ref{fig:hpmse}} shows the mean energy  of the highpass band
$\HP_{1}$ for the simulated compensation methods. For a compensated
wavelet transform, the highpass band can be regarded as the prediction
error. The mean energy is plotted against the refinement iterations
of the \meshbased{} compensation methods. The dotted blue curve results
from a traditional wavelet transform when no compensation is used
and thus it is independent of the iteration. The plot shows that the
energy in the highpass band can be reduced significantly by using
a compensation method within the wavelet transform. The dashed green
curve shows the result using a \zweid{} mesh. The comparison with
the results from the \dreid{} mesh method shows that a grid size
of 4 in $z$-direction is necessary to achieve better results than
the \zweid{} mesh. On the one hand, using a \dreid{} mesh has the
advantage that a displacement in $z$-direction can be compensated.
On the other hand, this leads to a subsampling of the motion information
in $z$-direction. 
\begin{table*}
\begin{centering}

\begin{tabular}{c|r|cc|ccc|ccc}

\toprule

Compensation method & \#mv~params & Mean energy & PSNR $\LP$ & \multicolumn{3}{c|}{\speckthreed{}
{[}MByte{]}} & \multicolumn{3}{c}{\jpthreed{} {[}MByte{]}}\tabularnewline

 &  & $\HP$ & {[}dB{]} & $\LP$ & $\HP$ & $\LP+\HP$ & $\LP$ & $\HP$ & $\LP+\HP$ \tabularnewline

\cmidrule{1-10}

none  &  0  & 2976.2 & 44.30 & 80.3 & 82.9 & 163.2 & 82.5 & 85.4 & 167.9 \tabularnewline

\addlinespace[1mm]

\zweid{} mesh, $16\times16$ &  261 888  & 1280.5 & 47.95 & 91.2 & 89.4 & 180.6 & 93.9 & 92.1 & 186.0
 \tabularnewline

\addlinespace[1mm]

\dreid{} mesh, $16\times16\times16$ &  26 037  & 1698.1 & 46.81 & 83.4 & 83.7 & 167.1 & 85.8 & 86.3 & 172.1 \tabularnewline

\dreid{} mesh, $16\times16\times8$ &  51 117  & 1422.7 & 47.58 & 84.0 & 84.1 & 168.1 & 86.4 & 86.7 & 173.1
 \tabularnewline

\dreid{} mesh, $16\times16\times4$ &  101 277  & 1232.8 & 48.23 & 84.7 & 84.8 & 169.5 & 87.1 & 87.5 & 174.6
 \tabularnewline

\addlinespace[1mm]

$\Delta$: \zweid{} $16\hspace{-1mm}\times\hspace{-1mm}16$ to \dreid{}
$16\hspace{-1mm}\times\hspace{-1mm}16\hspace{-1mm}\times\hspace{-1mm}4$
 & -61\% & -47.7 & +0.28 & -7.1\% & -5.1\% & -6.1\% & -7.2\% & -5.0\% & -6.1\%\tabularnewline

\bottomrule

\end{tabular} 

\end{centering}

\spaceBeforeLabel{}\spaceBeforeLabel{}

\caption{\label{tab:results}The table lists summarized results for the considered
compensation methods. \emph{\#mv params} denotes the number of parameters
needed for the model of the compensation method. The last row provides
a delta between the \zweid{} mesh $16\times16$ and the \dreid{}
mesh $16\times16\times4$}

\spaceBeforeLabel{}
\end{table*}
\begin{figure*}[!t]
\includegraphics[height=0.195\textwidth]{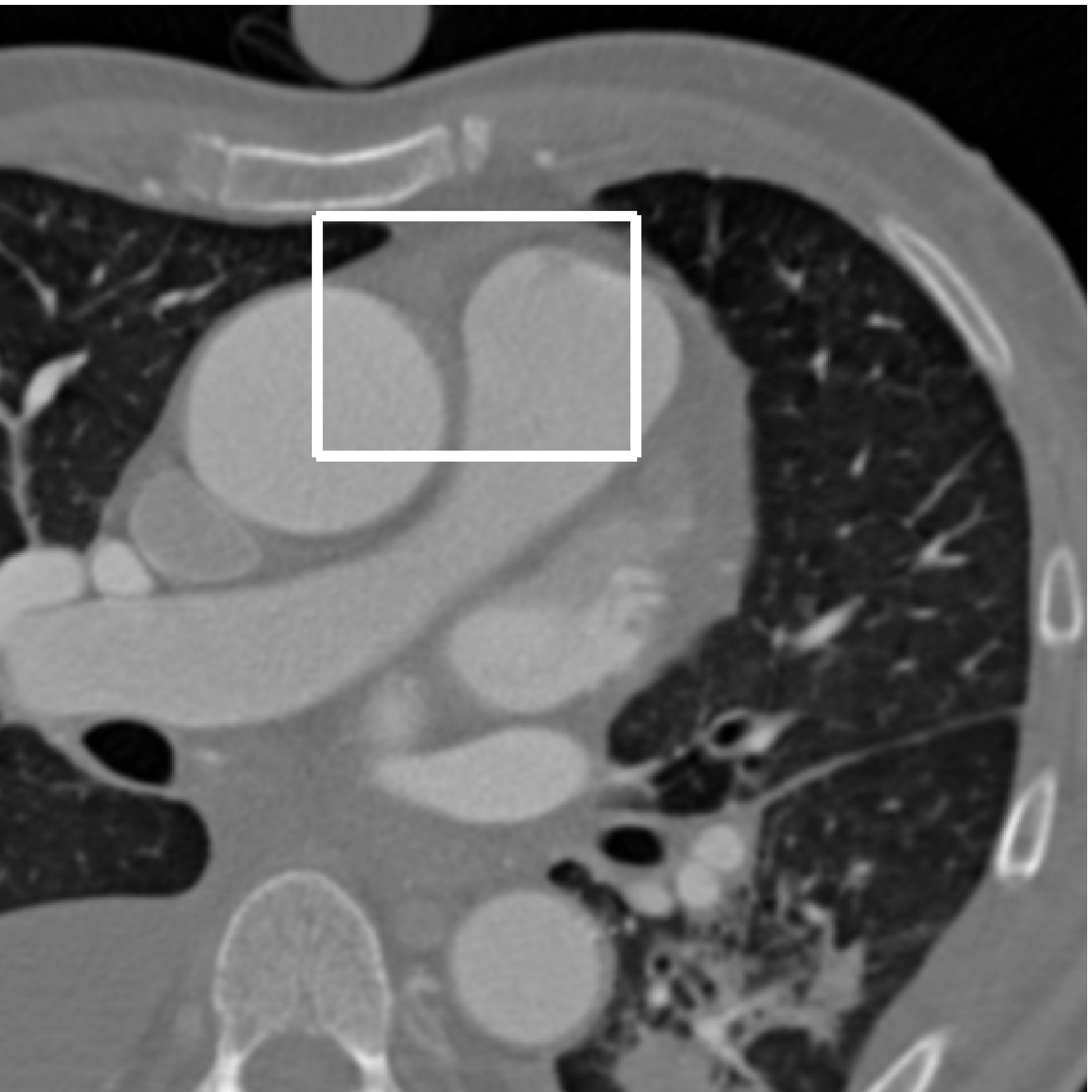}\hfill{}\includegraphics[height=0.195\textwidth]{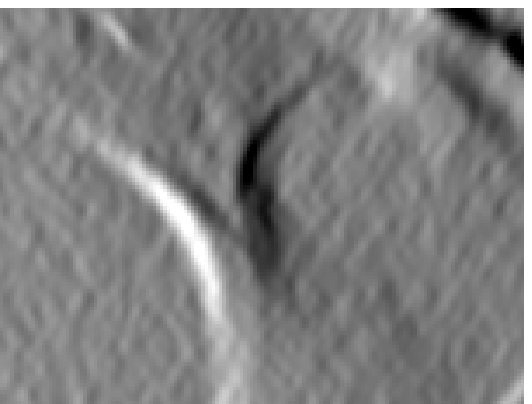}\hfill{}\includegraphics[height=0.195\textwidth]{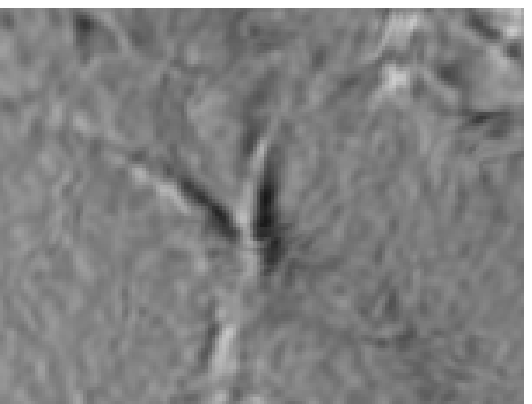}\hfill{}\includegraphics[height=0.195\textwidth]{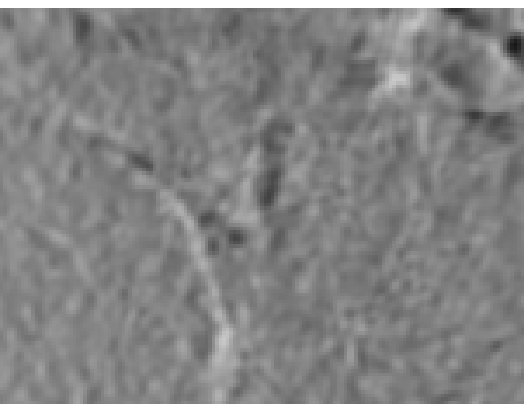}

original, $t=1$ \rotatebox[origin=c]{270}{$\Lsh$}, $t=2$ \rotatebox[origin=c]{270}{$\Rsh$}\hfill{}no
compensation, $\HP$ \rotatebox[origin=c]{270}{$\Lsh$}, $\LP$ \rotatebox[origin=c]{270}{$\Rsh$}\hfill{}\zweid{}
mesh $16\times16$, $\HP$ \rotatebox[origin=c]{270}{$\Lsh$}, $\LP$
\rotatebox[origin=c]{270}{$\Rsh$}\hfill{}\dreid{} mesh $16\times16\times4$,
$\HP$ \rotatebox[origin=c]{270}{$\Lsh$}, $\LP$ \rotatebox[origin=c]{270}{$\Rsh$}

\includegraphics[height=0.195\textwidth]{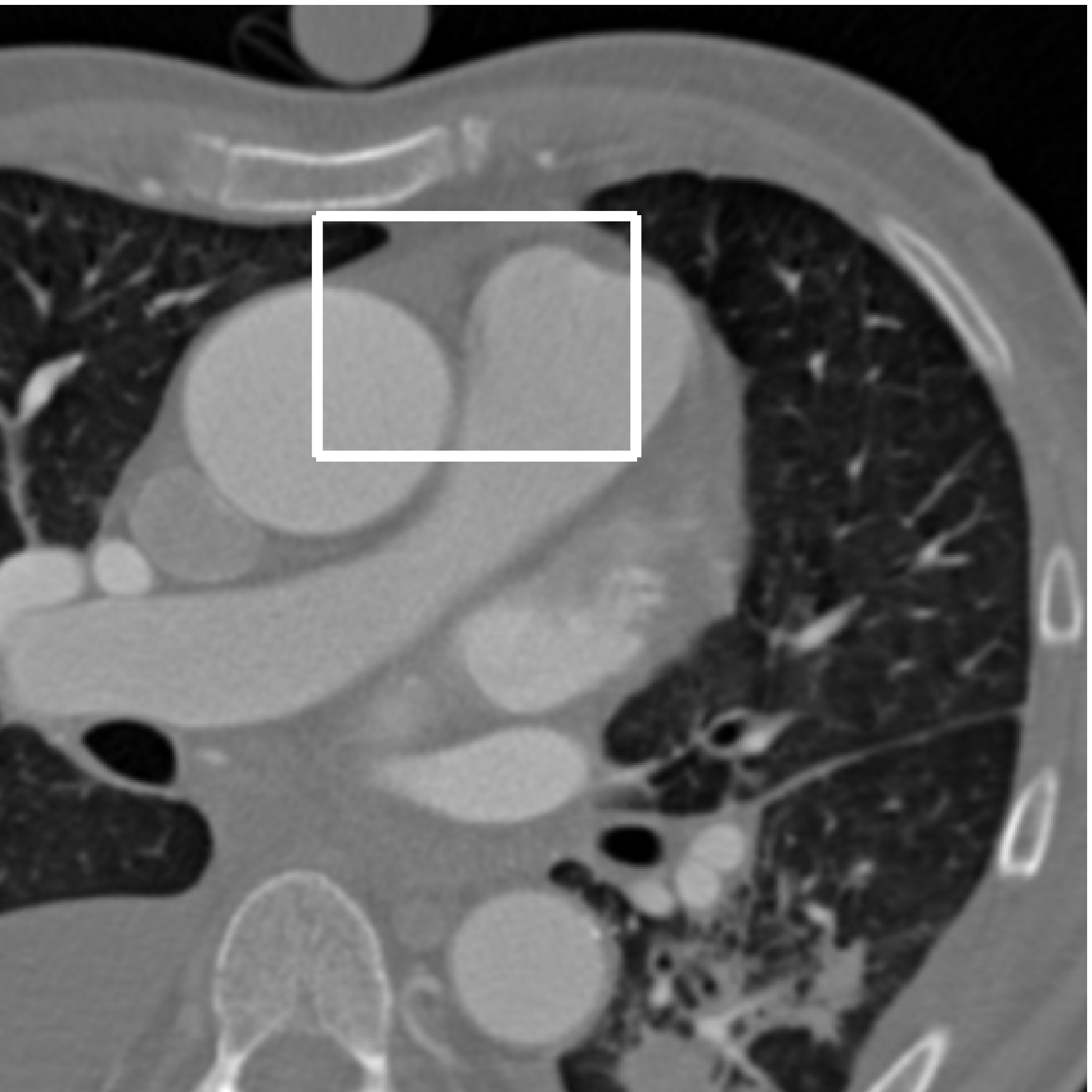}\hfill{}\includegraphics[height=0.195\textwidth]{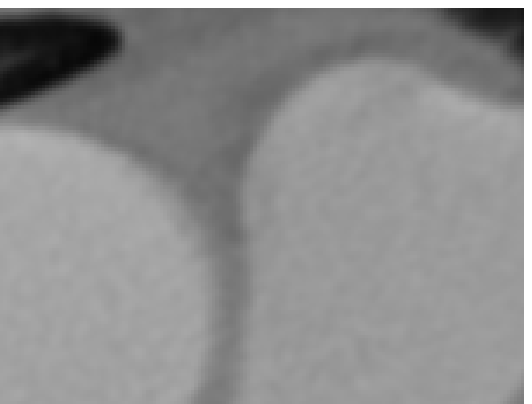}\hfill{}\includegraphics[height=0.195\textwidth]{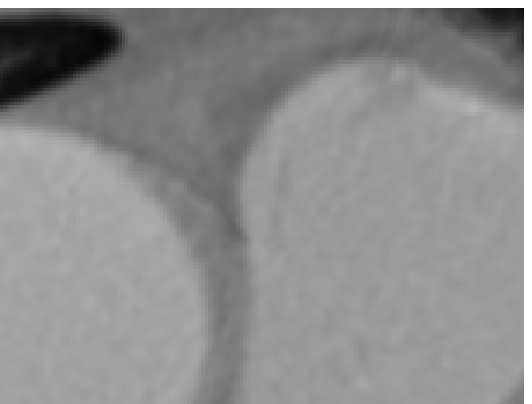}\hfill{}\includegraphics[height=0.195\textwidth]{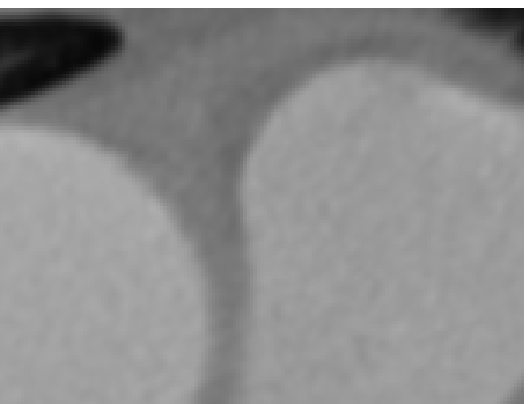}

\spaceBeforeLabel\spaceBeforeLabel\spaceBeforeLabel

\caption{\label{fig:visual-results}The first column shows temporal subsequent
images of the original volume at one slice position. The white areas
mark the origin of the details depicted in the remaining columns.
Details from the highpass bands ($\HP$, gray=0) are shown in the
top row and details from the lowpass bands ($\LP$) are shown in the
bottom row. The respective compensation methods are listed between.}

\spaceBeforeLabel
\end{figure*}

\Fig{\ref{fig:lppsnr}} shows the resulting quality of the lowpass
band in terms of $\text{PSNR}\left(\LP_{1},f_{1}\right)$ in dB. Again,
the metric is plotted against the iterations. The safety boundary
avoids the degeneration of the mesh structure. Although an approximation
is used for the inversion of the \meshbased{} methods, the quality
of the lowpass band is increased by nearly 5~dB compared to the traditional
transform. Further, the PSNR increases with a larger number of iterations.
Comparing the results from the \zweid{} mesh and the \dreid{} mesh
shows again, that for a grid size of 4 in $z$-direction, the \dreid{}
mesh yields better results. After 50 iterations, the \dreid{} mesh
leads to a PSNR gain of 0.15~dB compared to the \zweid{} mesh.

\Tab{\ref{tab:results}} summarizes the results for the wavelet transform
of all time steps of the \dreidt{} volume. The results are given
for 50 iterations, however, \Fig{\ref{fig:hpmse} and~\ref{fig:lppsnr}}
show that a convergence of the refinement of the mesh motion is reached
after about 20 iterations. The first column lists the compensation
methods used for the compensated wavelet transform applied in temporal
direction of the \dreidt{} volume. The second column lists the number
of parameters needed for the model of the compensation method. Although
a motion vector of the \dreid{} mesh has three components instead
of two, the overall number is smaller than for the \zweid{} mesh
due to the subsampling in $z$-direction. For the \dreid{} mesh with
grid size $16\times16\times4$, less than 40\% of the motion parameters
are needed compared to the \zweid{} mesh with grid size $16\times16$.
We assume that this leads to less side information for the \dreid{}
case but we do not consider coding of the motion information in the
following. The column MSE $\HP$ shows the mean energy of the highpass
band and the column PSNR $\LP$ shows the quality of the lowpass band
w.r.t.\  the corresponding reference volume $\sigf_{2t-1}$ after
the wavelet transform. The overall results are consistent with the
discussion of \Fig{\ref{fig:hpmse} and~\ref{fig:lppsnr}}.

To evaluate the compressibility, the resulting subbands were coded
losslessly using the standard wavelet coefficient coder set-partitioning
embedded block (\speckthreed{}) \cite{SPECKPatent}. We used the
implementation available for the QccPack library \cite{fowler2000qccpack}.
For comparison, the volumes were also coded using the wavelet-based
volume coder \jpthreed{} \cite{schelkens2006}. We used the OpenJPEG
\cite{openjpeg} implementation. For both subband volumes $\HP_{t}$
and $\LP_{t}$, further 2 wavelet decomposition steps in slice direction
and 5 wavelet decomposition steps in $xy$-direction were applied.
For \speckthreed{} we applied the \Legall{} as used in \jpthreed{}.

The columns entitled by \speckthreed{} and \jpthreed{} show the
respective results in MBbyte from lossless coding of the lowpass band,
the highpass band, and the sum of both.

Using \speckthreed{}, the results are a little better, but the overall
tendency is the same for both methods. When the quality of the lowpass
image is not of interest, no compensation method should be used when
lossless coding of a CT volume is considered because the data contains
a lot of correlated noisy structures that can be exploited by the
traditional wavelet transform without a compensation method. In contrast
to that, if a compensated wavelet transform is applied, it is not
possible to exploit the structures of the noise any more.

When the lowpass band is used as a scalable representation, the quality
is important. In this case, a compensation method can increase the
quality, as shown by the 4th column of \Tab{\ref{tab:results}}. The
coding results for the compensated transform suggest that for a \dreid{}
compensation it is better to use a \dreid{} \meshbased{} method
than  a \zweid{} \meshbased{} method. This might result from the
fact that the \dreid{} mesh compensation is not computed independently
slice-by-slice as for the \zweid{} mesh and thus leads to a smoother
prediction.

The visual examples depicted in \Fig{\ref{fig:visual-results}} support
the results from \Tab{\ref{tab:results}}. The lowpass band without
compensation is blurred. The \dreid{} mesh can exploit motion in
the $z$-direction over time and thus is able to further reduce the
energy in the highpass band resulting in a sharper lowpass with less
artifacts.

\section{Conclusion}

\label{sec:Conclusion}In this paper, we investigated \meshbased{}
compensation methods for a compensated wavelet transform of medical
\dreidt{} volumes that contain deforming displacements over time.
The proposed \dreid{} mesh compensation method is able to provide
a prediction for a complete volume and thus is able to exploit deforming
displacements in the third dimension as well. This is a huge advantage
compared to a slice-wise applied \zweid{} mesh compensation as proposed
in the literature.

The \dreid{} \meshbased{} compensation yields better compressible
subbands. Within our simulation data, the filesize of the lossless
coded subbands could be reduced by 6\% using a \dreid{} mesh instead
of a \zweid{} mesh. Further work aims at the investigation of a proper
inversion of the \meshbased{} compensation and a compensated transform
in $z$-direction within this framework.

\section*{Acknowledgment}

We gratefully acknowledge that this work has been supported by the
Deutsche Forschungsgemeinschaft (DFG) under contract number KA~926/4-2.

% References should be produced using the bibtex program from suitable% BiBTeX files (here: refs). The IEEEbib.bst bibliography% style file from IEEE produces unsorted bibliography list.% -------------------------------------------------------------------------

\bibliographystyle{IEEEbib}
\bibliography{bib/bibliography}

\begin{thebibliography}{10}

\bibitem{garbasTCSVT}
J.U. Garbas, B.~Pesquet-Popescu, and A.~Kaup,
\newblock ``{Methods and Tools for Wavelet-Based Scalable Multiview Video
  Coding},''
\newblock {\em {IEEE} Transactions on Circuits and Systems for Video
  Technology}, vol. 21, no. 2, pp. 113--126, Feb. 2011.

\bibitem{sanchez2010}
V.~Sanchez, R.~Abugharbieh, and P.~Nasiopoulos,
\newblock ``{3-D Scalable Medical Image Compression With Optimized Volume of
  Interest Coding},''
\newblock {\em {IEEE} Transactions on Medical Imaging}, vol. 29, no. 10, pp.
  1808--1820, Oct. 2010.

\bibitem{cavallaro2011region}
A.~Cavallaro, F.~Graf, H.-P. Kriegel, M.~Schubert, and M.~Thoma,
\newblock ``{Region of Interest Queries in CT Scans},''
\newblock in {\em Advances in Spatial and Temporal Databases}, pp. 56--73.
  Springer, 2011.

\bibitem{schnurrer2012vcip}
W.~Schnurrer, J.~Seiler, and A.~Kaup,
\newblock ``{Analysis of Displacement Compensation Methods for Wavelet Lifting
  of Medical 3-D Thorax CT Volume Data},''
\newblock in {\em Proceedings Visual Communications and Image Processing
  ({VCIP})}, San Diego, CA, USA, Nov. 2012, pp. 1--6.

\bibitem{schnurrer2012mmsp}
W.~Schnurrer, T.~Richter, J.~Seiler, and A.~Kaup,
\newblock ``{Analysis of Mesh-Based Motion Compensation in Wavelet Lifting of
  Dynamical 3-D+t CT Data},''
\newblock in {\em Proceedings {IEEE} International Workshop on Multimedia
  Signal Processing (MMSP)}, Banff, Canada, Sept. 2012, pp. 152--157.

\bibitem{secker2002}
A.~Secker and D.~Taubman,
\newblock ``{Highly Scalable Video Compression Using a Lifting-Based 3D Wavelet
  Transform with Deformable Mesh Motion Compensation},''
\newblock in {\em Proceedings International Conference on Image Processing
  ({ICIP})}, Rochester, NY, USA, June 2002, vol.~3, pp. 749--752.

\bibitem{calderbank1997}
A.R. Calderbank, I.~Daubechies, W.~Sweldens, and B.-L. Yeo,
\newblock ``{Lossless Image Compression Using Integer to Integer Wavelet
  Transforms},''
\newblock in {\em Proceedings International Conference on Image Processing
  ({ICIP})}, Washington, DC, USA, Oct. 1997, pp. 596--599.

\bibitem{nakaya1994}
Y.~Nakaya and H.~Harashima,
\newblock ``{Motion Compensation Based on Spatial Transformations},''
\newblock {\em {IEEE} Transactions on Circuits and Systems for Video
  Technology}, vol. 4, no. 3, pp. 339--356, June 1994.

\bibitem{Weinlich2012}
A.~Weinlich, P.~Amon, A.~Hutter, and A.~Kaup,
\newblock ``{Representation of Deformable Motion for Dynamic Cardiac Image Data
  Compression},''
\newblock in {\em Proceedings {SPIE} Medical Imaging}, San Diego, CA, USA, Feb.
  2012.

\bibitem{sullivan1991}
G.J. Sullivan and R.L. Baker,
\newblock ``{Motion Compensation for Video Compression Using Control Grid
  Interpolation},''
\newblock in {\em Proceedings {IEEE} International Conference on Acoustics,
  Speech, and Signal Processing ({ICASSP})}, Toronto, Canada, Apr. 1991, pp.
  2713--2716.

\bibitem{wang2012}
J.~Wang, M.~Yu, Y.~Xia, J.~Chen, and Z.~Xia,
\newblock ``{A Novel Method of DT Mesh Motion Estimation and Compensation based
  on RDWT},''
\newblock in {\em Proceedings {IEEE} International Conference on Communication
  Technology ({ICCT})}, Chengdu, China, Nov. 2012, pp. 1291--1295.

\bibitem{wang1996p1}
Y.~Wang and O.~Lee,
\newblock ``{Use of two-dimensional deformable mesh structures for video
  coding. Part I - The synthesis problem: mesh-based function approximation and
  mapping},''
\newblock {\em {IEEE} Transactions on Circuits and Systems for Video
  Technology}, vol. 6, no. 6, pp. 636 --646, Dec. 1996.

\bibitem{ohm2004interframe}
J.-R. Ohm, M.~Schaar, and J.W. Woods,
\newblock ``{Interframe Wavelet Coding - Motion Picture Representation for
  Universal Scalability},''
\newblock {\em Signal Processing: Image Communication}, vol. 19, no. 9, pp.
  877--908, 2004.

\bibitem{girod2005}
B.~Girod and S.~Han,
\newblock ``{Optimum Update for Motion-Compensated Lifting},''
\newblock {\em {IEEE} Signal Processing Letters}, vol. 12, no. 2, pp. 150--153,
  Feb. 2005.

\bibitem{schnurrer2013}
W.~Schnurrer, J.~Seiler, and A.~Kaup,
\newblock ``{Improving Block-Based Compensated Wavelet Lifting by
  Reconstructing Unconnected Pixels},''
\newblock in {\em Proceedings International Symposium on Signals, Circuits and
  Systems (ISSCS)}, Iasi, Romania, July 2013, pp. 1--4.

\bibitem{SPECKPatent}
W.~A. Pearlman and A.~Islam,
\newblock ``{Embedded and Efficient Low-complexity Hierarchical Image Coder and
  Corresponding Methods Therefor},'' Dec. 2003.

\bibitem{fowler2000qccpack}
J.E. Fowler,
\newblock ``{QccPack: An Open-Source Software Library for Quantization,
  Compression, and Coding},''
\newblock in {\em Proceedings Applications of Digital Image Processing XXIII},
  San Diego, CA, USA, Aug. 2000, vol. 4115, pp. 294--301.

\bibitem{schelkens2006}
P.~Schelkens, A.~Munteanu, A.~Tzannes, and C.~Brislawn,
\newblock ``{JPEG2000. Part 10. Volumetric data encoding},''
\newblock in {\em Proceedings {IEEE} International Symposium on Circuits and
  Systems ({ISCAS})}, Island of Kos, Greece, May 2006, pp. 3874--3877.

\bibitem{openjpeg}
A.~Descampe, F.~Devaux, H.~Drolon, D.~Janssens, and Y.~Verschueren,
\newblock ``{OpenJPEG~2.0.0},'' http://www.openjpeg.org, Nov. 2012.

\end{thebibliography}

\end{document}